\documentclass[twocolumn,twoside,slac,floatfix]{revtex4}

\usepackage{graphicx}
\usepackage{makeidx}
\usepackage{fancyhdr}

\begin{document}

\title{Bertini intra-nuclear cascade implementation in Geant4}

\author{Aatos Heikkinen, Nikita Stepanov} 
\affiliation{Helsinki Institute of Physics, P.O. Box 64, FIN-00014
  University of Helsinki, Finland}


\author{Johannes Peter Wellisch}
\affiliation{CERN, Geneva, Switzerland}
\begin{abstract}

We present here a intra-nuclear cascade model implemented in Geant4 5.0. 
The cascade model is based on re-engineering of INUCL code.
Models included are Bertini intra-nuclear cascade model with exitons, pre-equilibrium model, nucleus explosion model, 
fission model, and evaporation model.
Intermediate energy nuclear reactions from 100~MeV to 3~GeV energy are treated for proton, neutron, pions, photon and nuclear isotopes.
We represent overview of the models, review results achieved from simulations and make comparisons with experimental data.

\end{abstract}

\maketitle

\thispagestyle{fancy}


\section{INTRODUCTION}

The intra-nuclear cascade model (INC) was first proposed by Serber in 1947 \cite{serber47}.  
He noticed that, in particle-nuclear collisions the deBroglie wave-length of the incident particle is 
comparable to or shorter than the average intra-nucleon distance.
Hence, the justification for describing the interactions in terms of particle-particle  collisions.

The INC has been successfully used in the Monte Carlo simulations at intermediate energy region 
since Goldberger made first calculations by hand in 1947 \cite{goldberger48}. 
First computer simulations were done by Metropolis et al. in 1958 \cite{metropolis58}. 
Standard methods in INC implementations were formed when Bertini published his results in 1968 \cite{bertini68}.
An important addition was exciton model introduced by Griffin in 1966 \cite{griffin66}. 

Our presentation describes the implementation of a Bertini INC model in the Geant4 hadronic physics frameworks \cite{geant4collaboration03}.
Geant4\cite{geant4} is a Monte Carlo particle detector simulation toolkit, having applications also in the medical field and space
sciences. 
Geant4 provides a flexible framework for modular implementation of
various kinds of hadronic interactions. 
Geant4 exploits advanced Software Engineering techniques and Object
Oriented technology to achieve the transparency of the physics
implementation and to this way provide the possibility of validating the
physics results more easily. 

At level-2, the hadronic model frameworks are based on concepts of physics
processes, cross-section and models for final state generation. While the process is a general concept, models
are allowed to have restrictions in process type, material, element
and energy range.  Several models can be utilized by one process class; for instance, a
process class for inelastic scattering can use distinct models for different energies.

Process classes utilize model classes to determine the
secondaries produced in the interaction and to calculate the momenta
of the particles. Here we preset a collection of such models providing medium-energy
intra-nuclear cascade treatment.

\section{GEANT4 CASCADE MODEL}

In inelastic particle-nucleus collisions a fast phase ($10^{-23} - 10^{-22} s$) of INC results in a highly exited nucleus, 
and is followed by fission and pre-equilibrium emission. 
A slower ($10^{-18} - 10^{-16} s$) compound nucleus phase follows with evaporation.
A Boltzman equation must be solved to treat the physical process of the collision in detail.
 
The intra-nuclear cascade model developed by Bertini \cite{bertini68, bertini69, bertini71} solves the Boltzmann equation on the average.
The model has been implemented in several codes such as HETC \cite{alsmiller90}. 
Our model is based on re-engineering of INUCL code \cite{titarenko99a}.
Models included are a Bertini intra-nuclear cascade model with exitons, a pre-equilibrium model, a simple nuclear explosion model, a fission model, and a evaporation model. 

The nuclear model consist of a three-region approximation to the continuously changing density distribution of nuclear matter within nuclei.
Relativistic kinematics is applied throughout the cascade.
The cascade is stopped when all the particles which can escape the nucleus, have done so. 
Then conformity with the energy conservation law is checked.

\subsection{Model limits}

Particles treated are nucleons, pions, photons, and nuclear isotopes.
A bullet particle can be a proton, neutron or pion.
The range of targets allowed is arbitrary.

The necessary condition of validity of the INC model is $\lambda_{B} / v << \tau_{c} << \Delta t$, 
where $\lambda_{B}$ is the deBroglie wavelength of the nucleons, 
$v$ is the average relative nucleon-nucleon velocity and $\Delta t$ is the time interval between collisions.
The physical foundation becomes approximate at energies less than about $200 MeV$, and there needs to be supplemented with a pre-equilibrium model.
Also, at energies higher than 5-10 GeV the INC picture breaks down.
Our implementation in Geant4 has been tested with bullet kinetic energy between 100~MeV and 3~GeV.

\subsection{Intra-nuclear cascade model}

The basic steps of the INC model are summarized below:

\begin{enumerate}
\item The spatial point, where the incident particle enters, is selected uniformly over the projected area of the nucleus.
\item Total, free particle-particle cross-sections and region-dependent nucleon densities are used to select the path length for the projectile particle.
\item The momentum of a struck nucleon, the type of reaction, and the four momentum of the reaction products are determined.
\item the exciton model is updated as the cascade proceeds.
\item If Pauli's exclusion principle allows and $E_{particle} > E_{cutoff}$ = 2~MeV, step (2) is performed to transport the products.
\end{enumerate}

\begin{figure}
  \includegraphics[width=80mm,keepaspectratio]{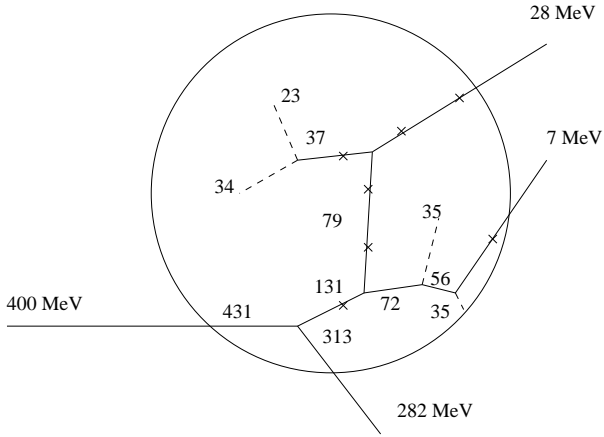}
  \caption{Schematic presentation of the intra-nuclear cascade. A hadron with 400 MeV energy is forming an INC history. Crosses present the Pauli exclusion principle in action. (The picture is a reproduction from original work of Bertini \cite{bertini68}.)}
  \label{figMC}
\end{figure}

After INC, the residual excitation energy of the resulting nucleus is used as input for a non-equilibrium model. The schematic presentation of Bertini's INC is shown in Fig.~\ref{figMC}

\subsection{The nuclear model}

Some of the basic features of the nuclear model are the following:

\begin{itemize}
\item The nucleons are assumed to to have a Fermi gas momentum distribution. 
The Fermi energy is calculated in a local density approximation i.e. 
it is made radius dependent with Fermi momentum $p_{F}(r) = (\frac{3 \pi^2 \rho(r)}{2})^\frac{1}{3}$.
\item The nucleon's binding energies (BE) are calculated using a mass formula.
A parametrization of the nuclear binding energy uses a combination of the Kummel mass formula, and
experimental data. Also, a asymptotic high temperature mass formula is used, 
for cases where it is not possible to use experimental data.
\end{itemize}

\subsubsection{Initialization}
The initialization phase fixes the nucleus radius  and momentum according to the Fermi gas model.

If the target is Hydrogen (A = 1), a direct particle-particle collision is performed, and no nuclear modeling is used.

If $1 < A < 4$, a nuclei model consisting one layer with radius of 8.0 fm is created.

For $4 < A < 11$, a nuclei model is composed of three concentric spheres $i = \{1, 2, 3\}$ with radii
$$r_{i}(\alpha_{i}) = \sqrt{C_{1}^{2} (1 - \frac{1}{A}) + 6.4} \sqrt{-log( \alpha_{i})}$$

where $\alpha_{i} = \{0.01, 0.3, 0.7\}$ and $C_{1} = 3.3836 A^{1/3}$

If $A > 11$, nuclei are modeled with three concentric spheres as well. The sphere radii are then defined as:
$$r_{i}(\alpha_{i}) =  C_{2} \log({\frac{1 + e^{- \frac{C_{1}}{C_{2}}}}{\alpha_{i}} - 1}) + C_{1}$$
where $C_{2} = 1.7234$.

The potential energy for nucleon N is
$$ V_{N} = \frac{p_{F}^2}{2 m_{N}} + BE_{N}(A, Z)$$
where $p_F$ is a Fermi momentum and BE a binding energy. 
 
Impulse distribution in each region follows Fermi distribution with zero temperature.

\begin{equation}
 f(p) = c p ^2
 \end{equation}

 where

 \begin{equation}
 \int_0^{p_F} f(p) dp = n_{p} \hspace{0.2truecm}  or \hspace{0.2truecm}   n_{n}.
 \end{equation}

 Here $n_p$ and $n_n$ are the number of protons or neutrons in a region and 
$p_F$ is momentum corresponding the Fermi energy

 \begin{equation}
 E_F = \frac{p_F^2}{2 m_N} = \frac{\hbar^2}{2 m_N}(\frac{3 \pi^{2}}{v})^\frac{2}{3},
 \end{equation}
 
 which depends on the density $n/v$ of particles, 
 and which is different for each particle and each region. 

\subsubsection{Pauli's exclusion principle}
Pauli's exclusion principle forbids interactions where the products would be in occupied states.
Following the assumption of a completely degenerate Fermi gas, the levels are filled from the lowest level.
The minimum energy allowed for a collision product corresponds to the lowest unfilled level of system, which is the Fermi energy in the region. 
So in practice, Pauli exclusion principle is taken into account by accepting only secondary nucleons which have $E_N > E_F$.

\subsubsection{Cross-sections and kinematics}

Path lengths of nucleons in the nucleus are sampled according to the local density and free nucleon-nucleon cross-sections.
Angles after collisions are sampled from experimental differential cross-sections.
Tabulated total reaction cross-sections are calculated by Letaw's formulation \cite{letaw83, letaw93, pearlstein89}.
For nucleon-nucleon cross-sections, parameterizations based on the experimental energy and isospin dependent data. 
The parameterization described in \cite{barashenkov72} is used. 

For pion the INC cross-sections are provided to treat elastic collisions, and inelastic channels:
$\pi^{-}$n $\rightarrow$ $\pi^{0}$n, $\pi^{0}$p $\rightarrow$ $\pi^{+}$n and $\pi^{0}$n $\rightarrow$ $\pi^{-}$p.
Multiple particle production is also implemented.

The S-wave pion absorption channels 
$\pi^{+}$nn $\rightarrow$ pn, $\pi^{+}$pn $\rightarrow$ pp, 
$\pi^{0}$nn $\rightarrow$ X , $\pi^{0}$pn $\rightarrow$ pn,      $\pi^{0}$pp $\rightarrow$ pp, 
$\pi^{-}$nn $\rightarrow$ X , $\pi^{-}$pn $\rightarrow$ nn , and $\pi^{-}$pp $\rightarrow$ pn are implemented.

\subsection{Pre-equilibrium model}

The Geant4 cascade model implements the exciton model proposed by Griffin \cite{griffin66}.
In this model nuclear states are characterized by the number of exited particles and holes (the exitons).
INC collisions give rise to a sequence of states characterized by increasing exciton number, eventually leading to a equilibrated nucleus.
For practical implementation of the exciton model we use level density parameters from \cite{ribansky73} and the matrix elements from \cite{kalbach78}.

In the exciton model the possible selection rules for particle-hole configurations in the course of the cascade are:
$\Delta p = 0, \pm 1$  $\Delta h = 0, \pm 1$  $\Delta n = 0, \pm 2$,
where p is the number of particle, h is number of holes and n = p + h is the number of exitons. 

The cascade pre-equilibrium model uses target excitation data, and
exciton configurations for neutron and proton to produce the non-equilibrium evaporation.
The angular distribution is isotropic in the frame of rest of the exciton system.

The parameterizations of the level density used, are tabulated both with their A and Z dependence and including a high temperature 
behavior. The nuclear binding energy is using a smooth liquid high energy formula.


\subsection{Break-up models}

Fermi break-up is allowed only in some extreme cases, i.e. for light nuclei ($A < 12$ and  $3 (A - Z) < Z < 6$ ) and 
if $E_{excitation} > 3 E_{binding}$ .
A simple explosion model decays the nucleus into neutrons and protons and decreases exotic evaporation processes.

The fission model is a phenomenological model using potential minimization. 
Binding energy parametrization is used and some features of the fission statistical model are incorporated as in\cite{fong69}.

\subsection{Evaporation model}

The statistical theory for particle emission from exited nuclei remaining after INC was originally developed by Weisskopf \cite{weisskopf37}. 
This model assumes complete energy equilibration before particle emission, and re-equilibration of excitation energies between successive evaporation emissions. 
As a result, the angular distribution of emitted particles is isotropic.

The Geant4 evaporation model for cascade implementation adapts the widely used computational method developed by Dostrowski \cite{dostrovsky59, dostrovsky60}.
The emission of particles is computed until the excitation energy fall below a cutoff. 
If a light nucleus is highly exited, a Fermi break-up model is executed. In addition, fission is performed when the fission channel is open. 
The main chain of evaporation is followed until  $E_{excitation}$ falls below E$_{cutoff}$ = 0.1 MeV. 
The evaporation model ends with a $\gamma$ emission chain, which is followed until E$_{excitation}$ $<$ E$^{\gamma}_{cutoff}$ = 10$^{-15}$ MeV.



\section{IMPLEMENTATION}


The Bertini cascade model is implemented in the Geant4 hadronic physics framework. The Bertini cascade source code module is located in directory {\it Geant4/\-source/\-processes/\-hadronic/\-models/\-cascade/\-cascade}.
All the models are used collectively through interface method {\it Apply\-Yourself} defined in a class {\it G4Cascade\-Interface}.
A Geant4 track  ({\it G4Track}) and a nucleus ({\it G4Nucleus}) are given as a parameters.

\begin{figure}
  \includegraphics[width=80mm,keepaspectratio]{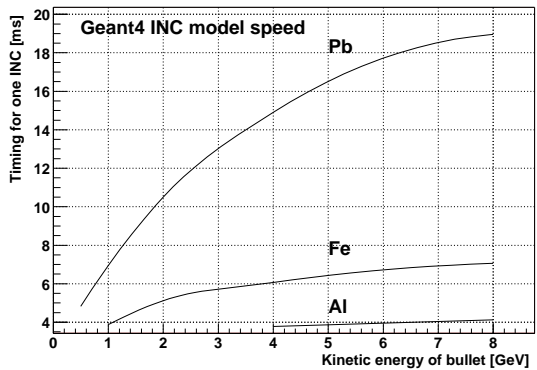}
  \caption{Speed of single INC performed for Al, Fe and Pb nuclei for proton bullet at energy-range 1-8 GeV.
A single PC with 300 MHz PII is used.}
  \label{timingModel}
\end{figure}

Currently implementation is not optimized for speed.
The typical timing behavior for different target nuclons and bullet energies is presented in Fig.~\ref{timingModel}.

\section{RESULTS}
\label{results}

We have tested the physics performance of Bertini cascade models in the first Geant4 5.0 implementation for proton induced reactions for energies from 100~MeV to 3~GeV. 
Detailed comparisons with experimental data has been made in energy range 160 -- 800 MeV.

\begin{figure}
  \includegraphics[width=80mm, keepaspectratio]{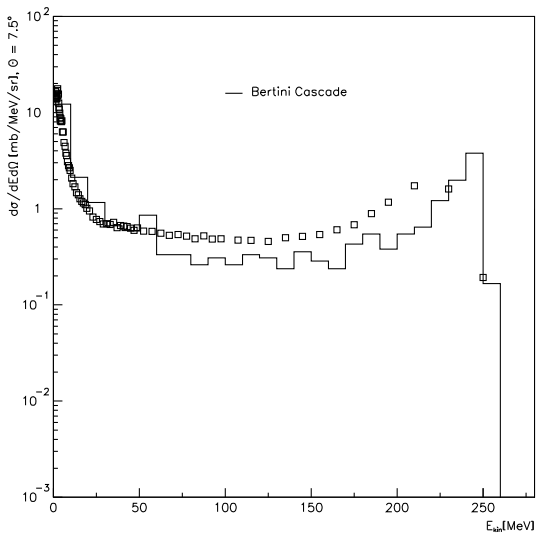}
  \caption{Geant4 Bertini INC model simulation of $p(256 MeV)$ + Fe $\rightarrow$ $n(\theta = 7.5^{\circ})$ + X in comparison with experimental data \cite{256iron}.}
  \label{n7}
\end{figure}

\begin{figure}
  \includegraphics[width=80mm,keepaspectratio]{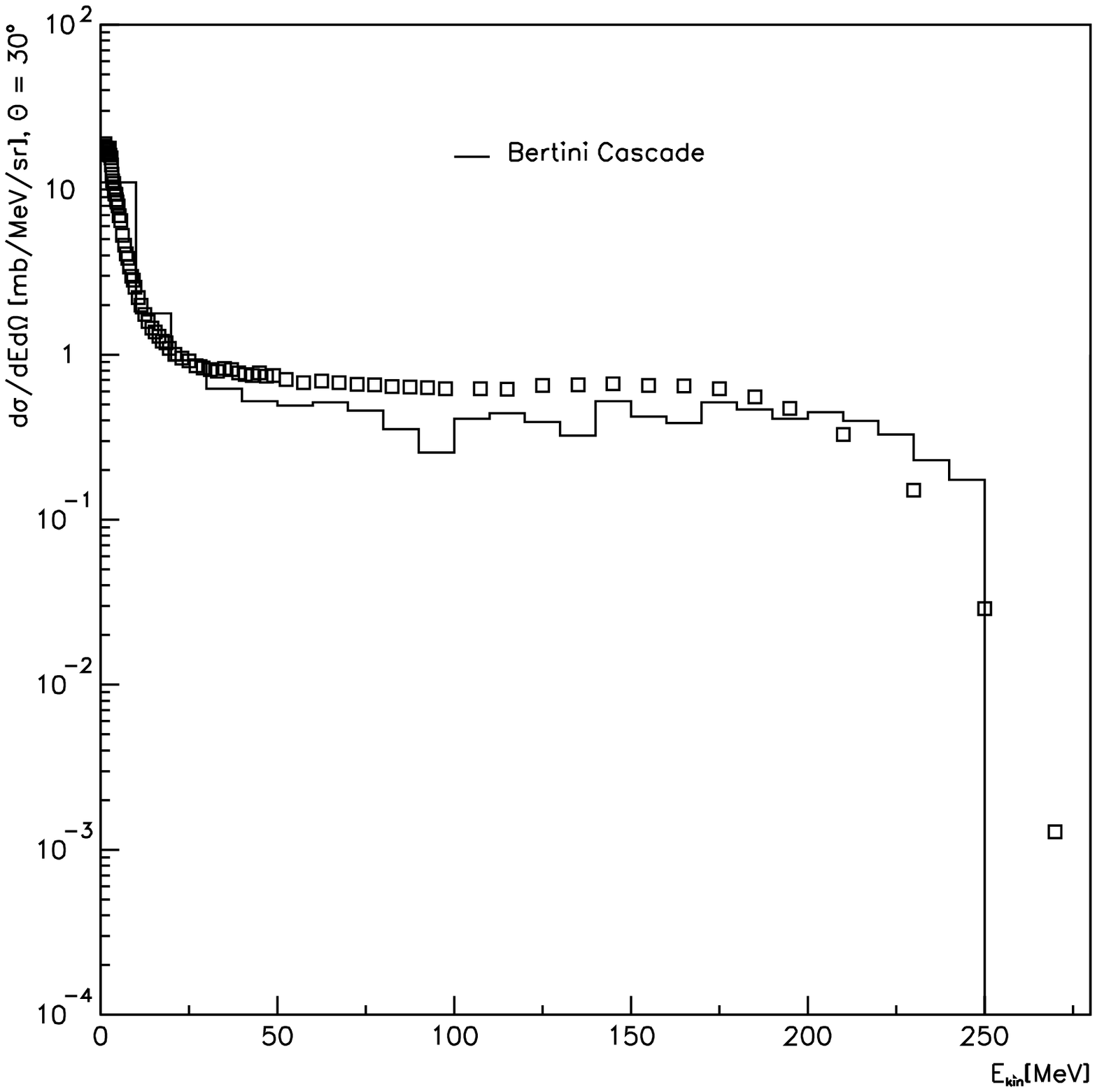}
  \caption{$p(256 MeV)$ + Fe $\rightarrow$ $n(\theta = 30^{\circ})$ + X.}
  \label{n30}
\end{figure}

\begin{figure}
  \includegraphics[width=80mm,keepaspectratio]{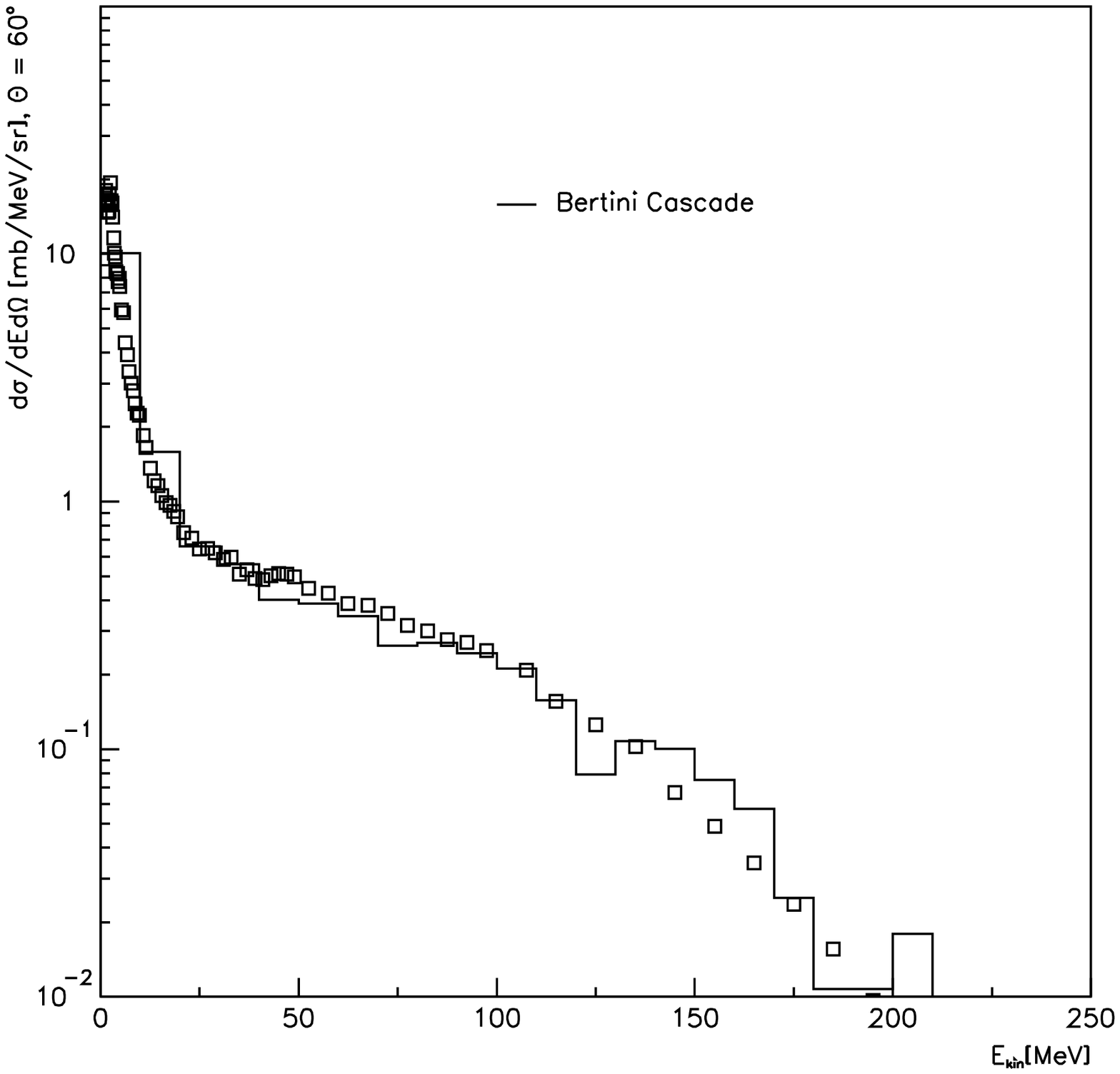}
  \caption{$p(256 MeV)$ + Fe $\rightarrow$ $n(\theta = 60^{\circ})$ + X.}
  \label{n60}
\end{figure}

\begin{figure}
  \includegraphics[width=80mm,keepaspectratio]{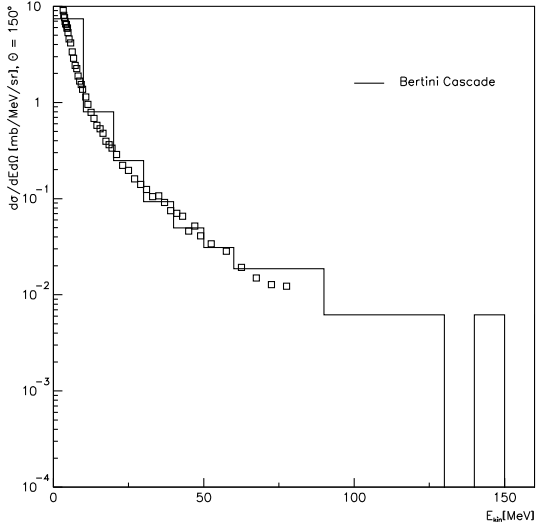}
  \caption{$p(256 MeV)$ + Fe $\rightarrow$ $n(\theta = 150^{\circ})$ + X.}
  \label{n150}
\end{figure}

The double differential cross-sections for neutrons at angles of $7.5^{\circ}$, $30^{\circ}$, $60^{\circ}$, and $150^{\circ}$  are presented in Figs.~\ref{n7}-\ref{n150}. Proton incident particle at 256~MeV energy is hitting an iron target and all Bertini sub-models are used.
Data were taken from \cite{256iron}.


\begin{figure}
  \includegraphics[width=80mm,keepaspectratio]{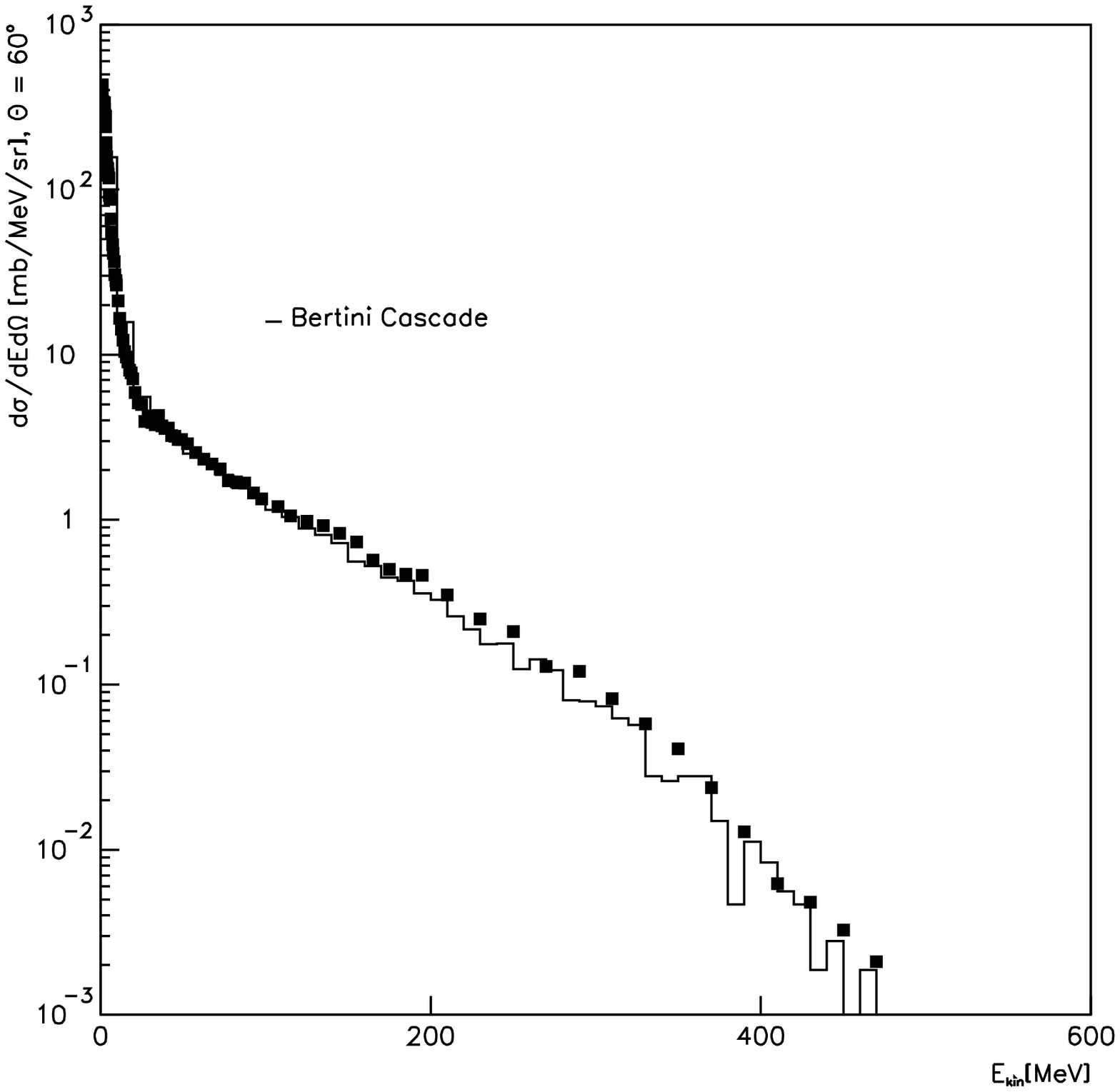}
  \caption{Geant4 Bertini INC model simulation of 
$p(597 MeV)$ + Pb $\rightarrow$ $n(\theta = 60^{\circ})$ + X 
in comparison with experimental data \cite{597lead}.}
  \label{pbn60}
\end{figure}

\begin{figure}
  \includegraphics[width=80mm,keepaspectratio]{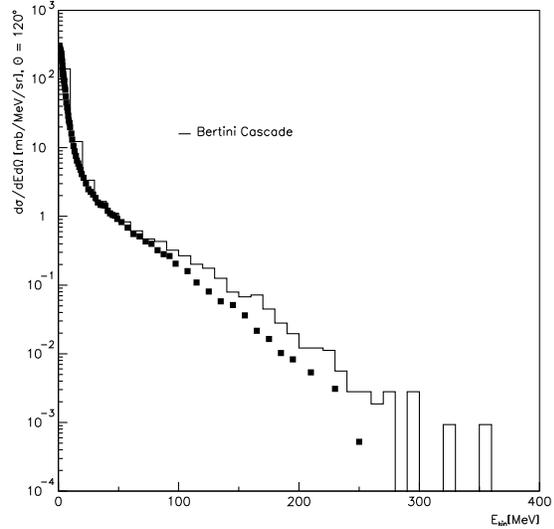}
  \caption{$p(597 MeV)$ + Pb $\rightarrow$ $n(\theta = 120^{\circ})$ + X.}
  \label{pbn120}
\end{figure}

Figures \ref{pbn60} and \ref{pbn120} show double differential cross-section of neutrons resulting from 597 MeV protons impinging on lead. Data were taken from \cite{597lead}

\begin{figure}
  \includegraphics[width=80mm,keepaspectratio]{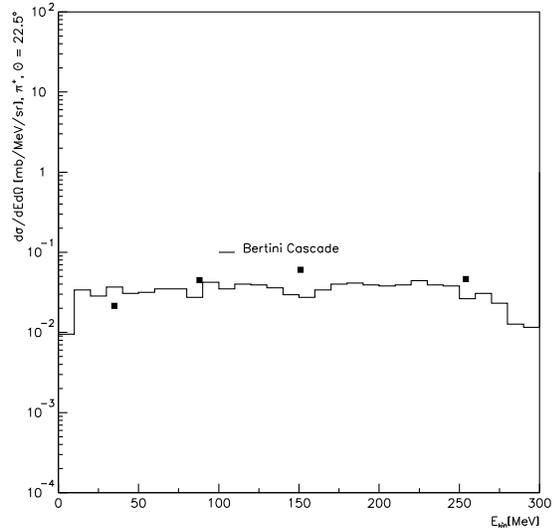}
  \caption{Geant4 Bertini INC model simulation of
$p(585 MeV)$ + Pb $\rightarrow$ $\pi^{+}(\theta = 22.5^{\circ})$ + X
in comparison with experimental data \cite{585lead}.}
  \label{pi22}
\end{figure}

\begin{figure}
  \includegraphics[width=80mm,keepaspectratio]{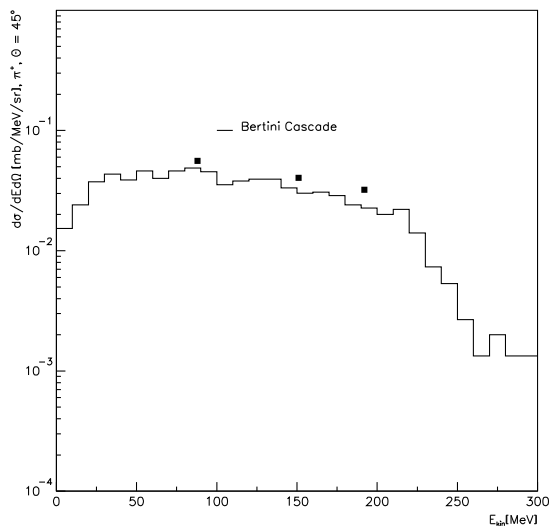}
  \caption{$p(585 MeV)$ + Pb $\rightarrow$ $\pi^{+}(\theta = 45^{\circ})$ + X.}
  \label{pi45}
\end{figure}

\begin{figure}
  \includegraphics[width=80mm,keepaspectratio]{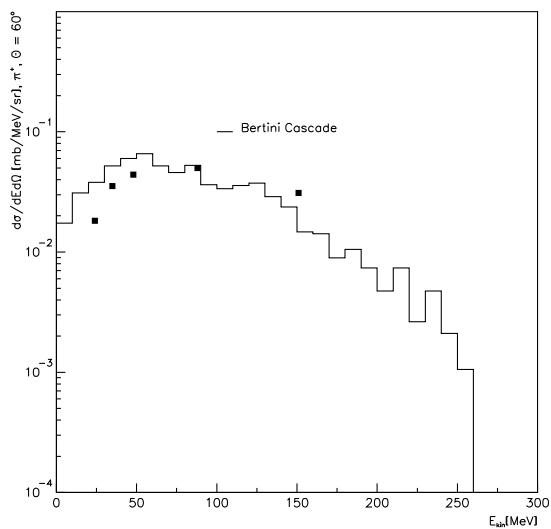}
  \caption{$p(585 MeV)$ + Pb $\rightarrow$ $\pi^{+}(\theta = 60^{\circ})$ + X.}
  \label{pi60}
\end{figure}

\begin{figure}
  \includegraphics[width=80mm,keepaspectratio]{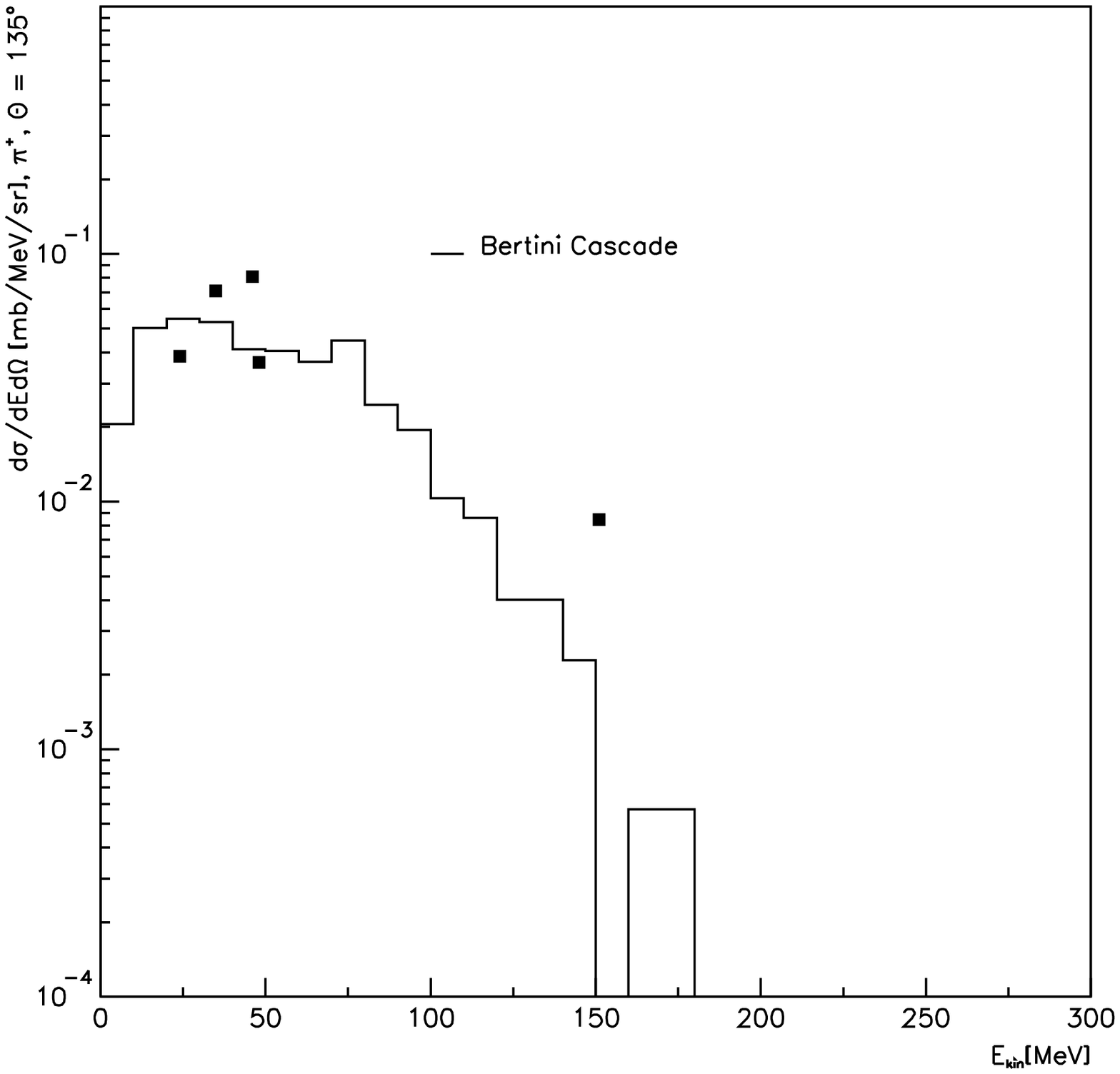}
  \caption{$p(585 MeV)$ + Pb $\rightarrow$ $\pi^{+}(\theta = 135^{\circ})$ + X.}
  \label{pi135}
\end{figure}

Finally, an example of pion production physics performance.
Double differential cross-sections for $\pi^{+}$ are given for
angels from $22.5^{\circ}$ to $135^{\circ}$ in Figs.~\ref{pi22}-\ref{pi135}.

The agreement with experimental data is found to be from moderate to relatively good. Data fo the 585~MeV pion comparisons were taken from \cite{585lead}.

\section{CONCLUSION}
We have released a Bertini INC model in Geant4.
Exitons, pre-equilibrium, nucleus explosion, fission, and evaporation are modeled.
Particles treated are $\gamma$, $\pi$, n, p, and nuclear isotopes, and incident nucleons and pions can be treated.
We have tested the code in a energy range 60~MeV - 3~GeV and found a reasonable agreement with
experimental data. Tuning of performance and models are planned for future Geant4 releases.
We also plan to extend the upper energy limit of the models.







\end{document}